# Thermoelectric Properties of Mg doped Mercury Selenide HgSe

Y. SELMANI [1, *], H. LABRIM [1,2], A. JABAR [3], L. BAHMAD [1]

[1] Laboratory of Condensed Matter and Interdisciplinary Sciences (LaMCScI), Faculty of Sciences, Mohammed V University, Av. Ibn Batouta, B. P. 1014 Rabat, Morocco

[2] Advanced Systems Engineering Laboratory, National School of Applied Sciences, Ibn Tofaîl University, Kenitra, Morocco

[3] LPMAT, Faculty of Sciences Aïn Chock, Hassan II University of Casablanca, B.P. 5366 Casablanca, Morocco

## Abstract:

Using the density functional theory (DFT) in combination with Boltzmann transport theory, the influence of Mg concentrations (x) doping on the thermoelectric properties of $Hg_{1-x}Mg_xSe$ ternary alloys was systematically investigated. The generalized gradient approximations of Perdew-Burke-Ernzerhof (GGA-PBE) have been used to illustrate the exchange correlation potential. Various thermoelectric transport parameters, such as the Seebeck coefficient (**S**), the thermal conductivity over relaxation time, the electrical conductivity over relaxation time, the power factor *(PF)* and the figure of merit *(ZT)* have been deduced and discussed. The obtained results of thermoelectric properties show that the studied materials can be useful for room temperature thermoelectric devices. It is also found that Mg compositions can increase the thermal efficiency of the HgSe alloy.

---

[1]) Corresponding author: ahmedselmani123@gmail.com (Y. SELMANI).

# 1. Introduction

The conversion of thermal energy into electricity necessitates specific thermoelectric materials. This kind of materials have attracted increasing interest in the scientist community [1-6]. The efficiency of thermoelectric energy conversion of these materials depends on the figure of merit *ZT*. This later is related directly to the Seebeck coefficient *S*, the electrical conductivity $\sigma$ and the thermal conductivity *k* [7]. However, the large values of the figure of merit *ZT* have been achieved in thermoelectric materials via carrier optimization [8-10] and band structure engineering [11-15]. The challenge to create high performance thermoelectric materials lies in achieving simultaneously large Seebeck coefficient, high electrical conductivity and low thermal conductivity.

Recently, various binary and ternary group of chalcogenide compounds were investigated such as $Bi_2Te_3$ [16-17], $Bi_2Se_3$ [18], $Sb_2Te_3$ [19-20], $Sb_2Se_3$ [21], $Sb_2Te_{3-x}Se_x$ [22], $Bi_2Se_xTe_{3-x}$ [23-24] and $Bi_{2-y}Sb_yTe_3$ [25-27]. All of these compounds are highly promising thermoelectric materials in the temperature range around the room temperature [28-28]. In addition, some material systems have been developed to be as high thermoelectric performance with ZT value beyond unity [29-32].

Moreover, semiconductors based on magnesium pnictides with general formula $Mg_3X_2$ (X= As, Sb, Bi) are widely studied due to their functional applications in the thermoelectric field [33–35]. Furthermore, various binary and ternary of $Mg_3X_2$-based Zintl compounds were investigated, such as $Mg_3Bi_2$ [36-37], $Mg_3Sb_2$ [38-39] and $Mg_3Sb_{2-x}Bi_x$ [40-43]. In addition, J. Xin *et al.* [44] have studied p-type $Mg_3X_2$ (X= Sb, Bi) single crystals by using a self-flux method and the Debye-Callaway model. They showed that the transport properties of such materials could pave a way to realize enhanced thermoelectric performance in single-crystalline $Mg_3X_2$-based Zintl compounds.

Previously, researchers have investigated the thermoelectric performance of semiconductors based on the magnesium chalcogenides such as magnesium selenide (MgSe) and magnesium telluride (MgTe). R. Muthaiah and their coworkers [45-46] have illustrated the thermal conductivity (k) of MgTe and MgSe compounds with various crystallographic phases such as rocksalt, zincblende, wurtzite and nickel arsenic (NiAs), using density functional theory and Boltzmann transport equation. They reported that the MgTe and MgSe materials shows a promising nature for thermoelectric applications.

The mercury selenide HgSe which crystallizes in zincblende structure is technologically interesting material due to their applications in optoelectronic devices. Mycielski *et al.* [47] band structure of HgSe from inter-band magneto-absorption spectra and found that material reveals as zero-gap semiconductor. The optical absorption spectra of HgSe were studied by Szuszkiewi and reveals its metallic character with band gap about −0.25 eV [48]. Cardona *et al.* [49] have carried out first principles calculations of electronic, vibrational, thermodynamic properties and phonon dispersion of zinc-blende mercury chalcogenide HgSe, and HgTe. The structural, electronic, optical and dynamic properties of MgSe binary compound have been performed by M. N. Secuk and coworkers [50]. Regarding the thermoelectric properties, the thermal conductivity values in semimetals materials could be small, especially if they consist of heavy elements. G. A. Slack [51] investigated the thermal conductivity and found that parameter value is about 1.7 Wm$^{-1}$K$^{-1}$ in HgSe at room temperature. Several works on different other aspects of mercury selenide HgSe, include electron mobility and thermoelectric power from electron scattering in HgSe [52-54]

In this study, we investigate the effect of Mg doping on the thermoelectric properties of semimetal HgSe compound by using the combined method of the first principles calculations and the semi-classical Boltzmann theory. The Seebeck coefficient (**S**), the thermal conductivity over relaxation (**k/τ**), the electrical conductivity over relaxation (**σ/τ**), the power factor *(PF)* and the figure of merit *(ZT)* are studied.

This paper is organized as follows: In section II, we present the details of calculations. In section III, we discuss the obtained results of the thermoelectric properties of the studied alloys. Finally, conclusions are summarized Section IV.

## 2. Computational method

Our calculations are based on density functional theory via the ABINIT package [55]. The generalized gradient approximations (GGA) [56] Perdew Burke Ernzerhof (PBE) functional [57] was employed to consider the interaction of the exchange-correlation (XC) function. Fritiz Haber Institute (FHI) pseudopotential files of each atom [58] were gotten from the ABINIT software web site [59], are used in the computation. The kinetic energy cutoff value of 435 eV, the k-point mesh of $8 \times 8 \times 8$ and theorical values of lattice constants were gotten from other studies [60], and they were used to obtain the DFT outputs. In fact, we combined the outputs of DFT to transport theory based on Boltzmann equation to calculate the transport properties. Moreover, the thermoelectric parameters have been calculated by using the semi-classical Boltzmann transport theory within the relaxation time *(τ)* approximations as performed in the BoltzTrap code [61].

## 3. Results and discussion

Based on the combination between the DFT method as implemented in ABINIT code and semi-classical Boltzmann theory as incorporated in BoltzTraP code we have calculate the thermoelectric properties using the following Boltzman semi classical expressions [62-64]:

$$\sigma_{\alpha,\beta}(\varepsilon) = \frac{e^2}{N}\sum_{i,k}\tau_{i,k}kv_{\alpha}(i,\vec{k})v_{\beta}(i,\vec{k})\delta(\varepsilon-\varepsilon_{i,k}) \qquad (1)$$

Were $\tau$ is the relaxation time that is taken as a constant in BoltzTraP code [61] (typical value around $10^{-14}$ s). While e represents charge of electron, N stands for the number of k-points, $v_{\alpha}(i,\vec{k})$ and $v_{\beta}(i,\vec{k})$ are a components of group velocities.

The thermoelectric coefficients such as the Seebeck coefficient $S_{\alpha,\beta}$, the electrical conductivity $\sigma_{\alpha,\beta}$, and the electronic thermal conductivity $k^0_{\alpha,\beta}$ tensors are investigated as a function of temperature T and chemical potential $\mu$, by integrating the transport distribution:

$$S_{\alpha,\beta}(T,\mu) = \frac{1}{eT\Omega\sigma_{\alpha,\beta}(T,\mu)}\int \sigma_{\alpha,\beta}(\varepsilon)(\varepsilon-\mu)\left[-\frac{\partial f_0(T,\mu,\varepsilon)}{\partial \varepsilon}\right]d\varepsilon \qquad (2)$$

$$k^0_{\alpha,\beta}(T,\mu) = \frac{1}{e^2T\Omega}\int \sigma_{\alpha,\beta}(\varepsilon)(\varepsilon-\mu)^2\left[-\frac{\partial f_0(T,\mu,\varepsilon)}{\partial \varepsilon}\right]d\varepsilon \qquad (3)$$

$$\sigma_{\alpha,\beta}(T,\mu) = \frac{1}{\Omega}\int \sigma_{\alpha,\beta}(\varepsilon)\left[-\frac{\partial f_0(T,\mu,\varepsilon)}{\partial \varepsilon}\right]d\varepsilon \qquad (4)$$

Here, $\Omega$ is the volume of the unit cell, α and β indicate the tensor indices, and $f_0$ denote Fermi–Dirac distribution function.

### 3.1. Seebeck coefficient

The Seebeck coefficient **S** measures the potential difference between two different conductors or semiconductors, when there is a temperature gradient between the two junctions. This parameter is important to describe the thermoelectric properties of such materials. The obtained results Seebeck coefficient for the studied materials are depicted as a function of temperature in Fig. 1. From this figure, the positive values of S indicate that the studied materials exhibit p-type conducting behavior. At low temperature, it is clearly observed that The Seebeck coefficient starts increases with increasing temperature to reaches important values at 300 K, which are 2.00, 2.18, 2.28, 2.10 and $2.12 \times 10^{-4}$ V/K for $Hg_{1-x}Mg_xSe$ (x=0.0, 0.25, 0.50, 0.75 and 1.0), respectively. Then the S curves decrease slowly up to 600 K for x = 1.0, while for other dopant concentrations × 0.0, 0.25 0.50 and 0.75, its values increase slowly. At high temperatures, up to 1200K, the Seebeck coefficient values is gradually decreases for all Mg concentration. We have also calculated the Seebeck coefficient as a function of Mg concentrations, see Fig. 2. From this figure, we can note that the S parameter increases by increasing the temperature. In addition, It can be seen from S curves that the Seebeck coefficient values increases as the Mg concentrations increases up to 0.50, and then it start to decreases.

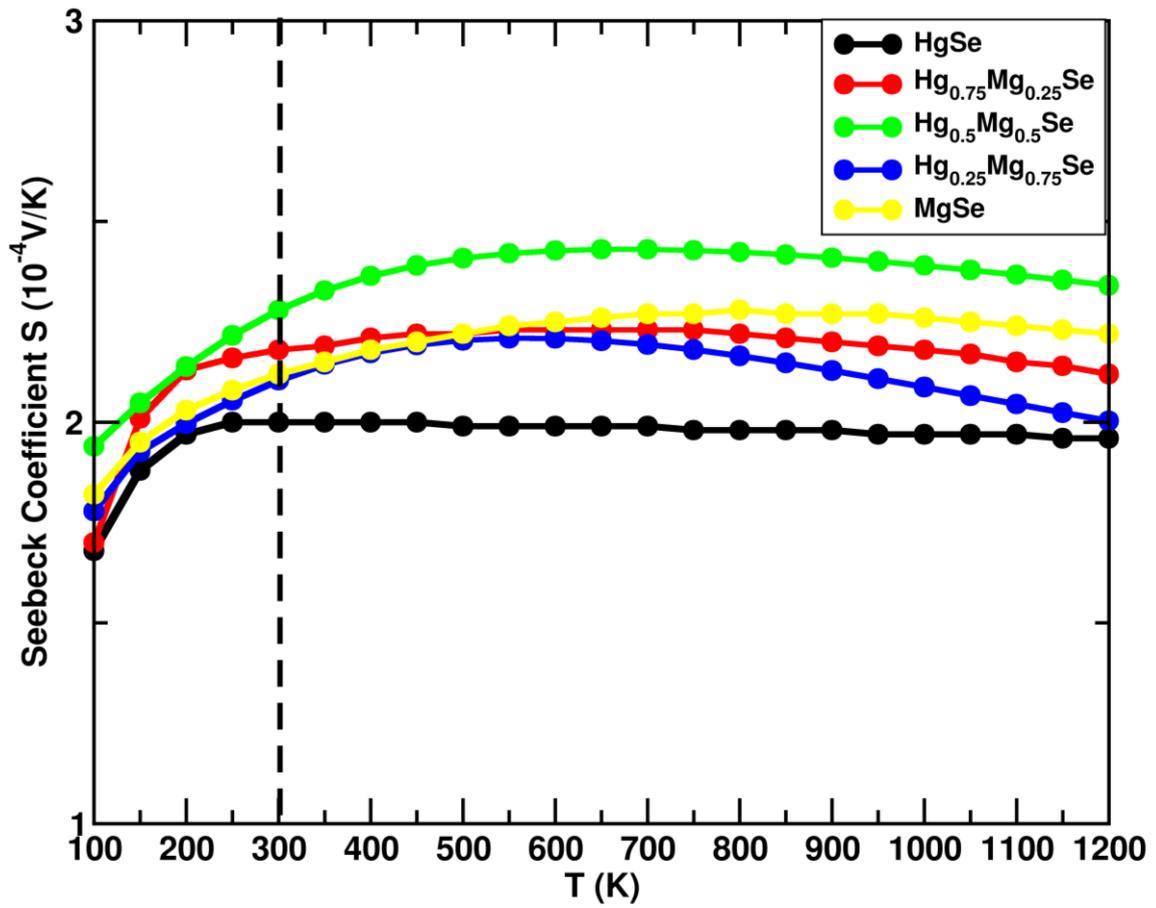

*Figure 1: Calculated Seebeck coefficient (S) versus temperature (T).*

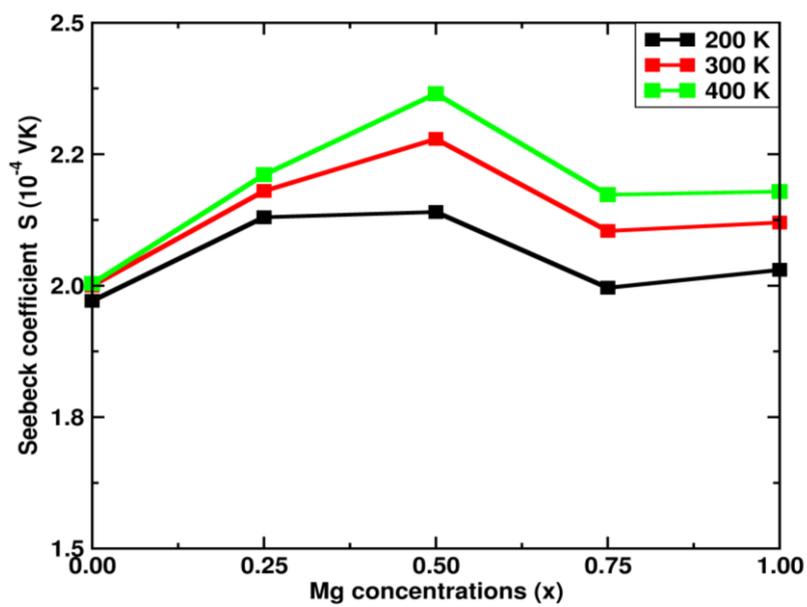

*Figure 2: Seebeck coefficient as a function of Mg concentrations for different temperature.*

## 3.2. Electronic Thermal conductivity

The electronic thermal conductivity **(k/τ)** is a physical quantity that characterizes the ability of a material to heat up. Fig. 3 displayed the total thermal conductivity of $Hg_{1-x}Mg_xSe$ (x=0.0, 0.25, 0.50, 0.75 and 1.0) as a function of temperature. Form this figure it is clearly observed that the **k/τ** started from 0 at low temperature, then it increases with increasing temperature until it reaches high values at high temperature. We also see that at a lower temperature than room temperature, all the studied materials have a low thermal conductivity less than $2 \times 10^{14}$ W/mKs. At a higher temperature, The MgSe has a higher thermal conductivity than other studied materials. Furthermore, with increasing the doping concentration of Hg, thermal conductivity first decreases at x = 0.25, then increases up x = 1.0 ($Hg_{1-x}Mg_xSe$). Fig. 4 shows the electronic thermal conductivity as a function of Mg concentrations. From this figure we noticed that the conductivity remains constant up to a concentration of 0.25. Then it begins to reach minimum values at x=0.50. Beyond 0.50 of the concentrations, it is found that the thermal conductivity with the increase in Mg concentrations.

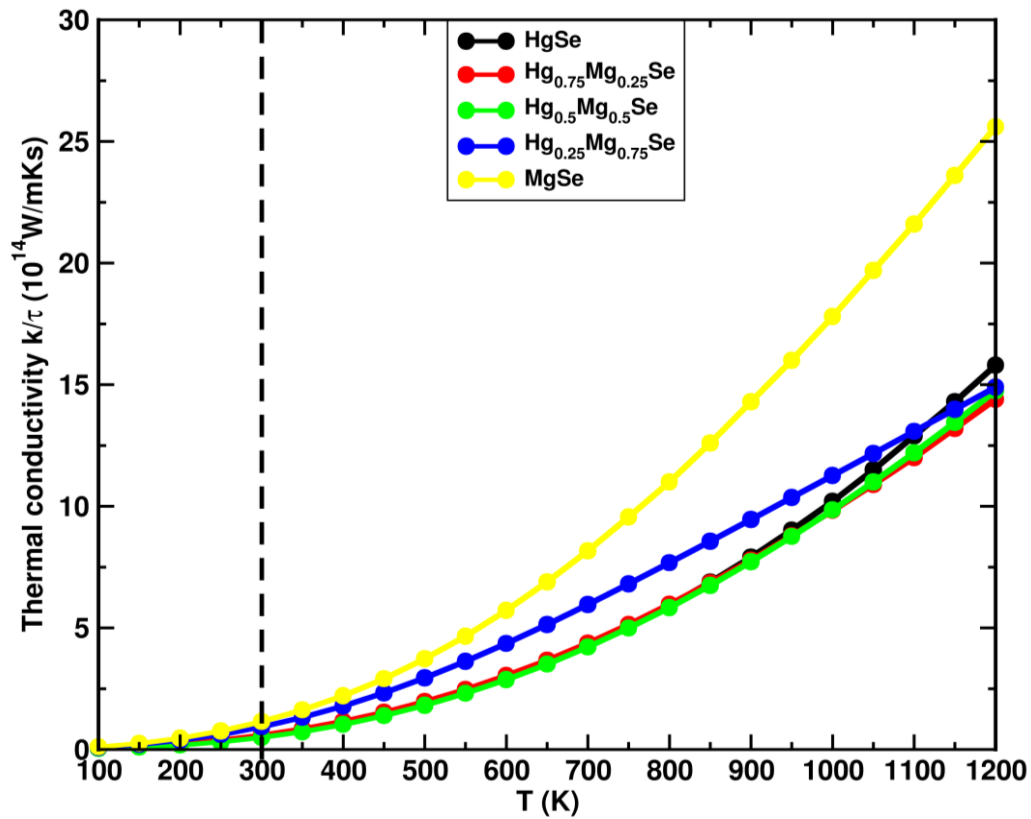

*Figure 3: Calculated electronic thermal conductivity over relaxation time (k/τ) versus temperature (T).*

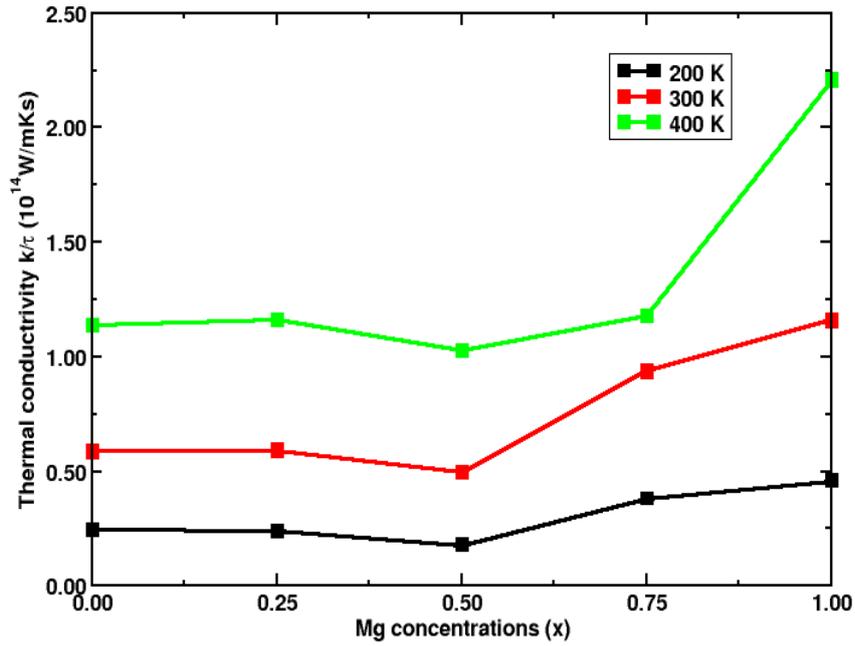

*Figure 4: Thermal conductivity versus Mg concentrations for different temperature.*

### 3.3. Electrical conductivity

The effect of temperature on the electrical conductivity ($\sigma/\tau$) for all studied materials are given in Fig. 5. Generally, the electrical conductivities in the studied materials showed similar curves to the electronic thermal conductivities. From $\sigma/\tau$ curves, it is clearly shows that the MgSe binary compound has a large electrical conductivity compared to the others of the studied materials. Additionally, at 300 K, the electrical conductivity values of $Hg_{1-x}Mg_xSe$ (x=0.0, 0.25, 0.50, 0.75 and 1.0) compounds are, respectively. When the temperature exceeds 300K, the electrical conductivity movement registers a noticeable rise. Moreover, the Mg concentrations cause an increase in electrical conductivity, which is good for thermoelectric applications. As shown in Fig. 6 the electrical conductivity increases with increasing the temperature. Also, it can be noted that when the x value increases up to 0.50, the electrical conductivity decreases until to reaches minimum values. Beyond the 0.50 of Mg concentrations, the electrical conductivity gradually increases. The obtained results of Seebeck coefficient, electronic thermal and electrical conductivity are summarized in the Table. 1.

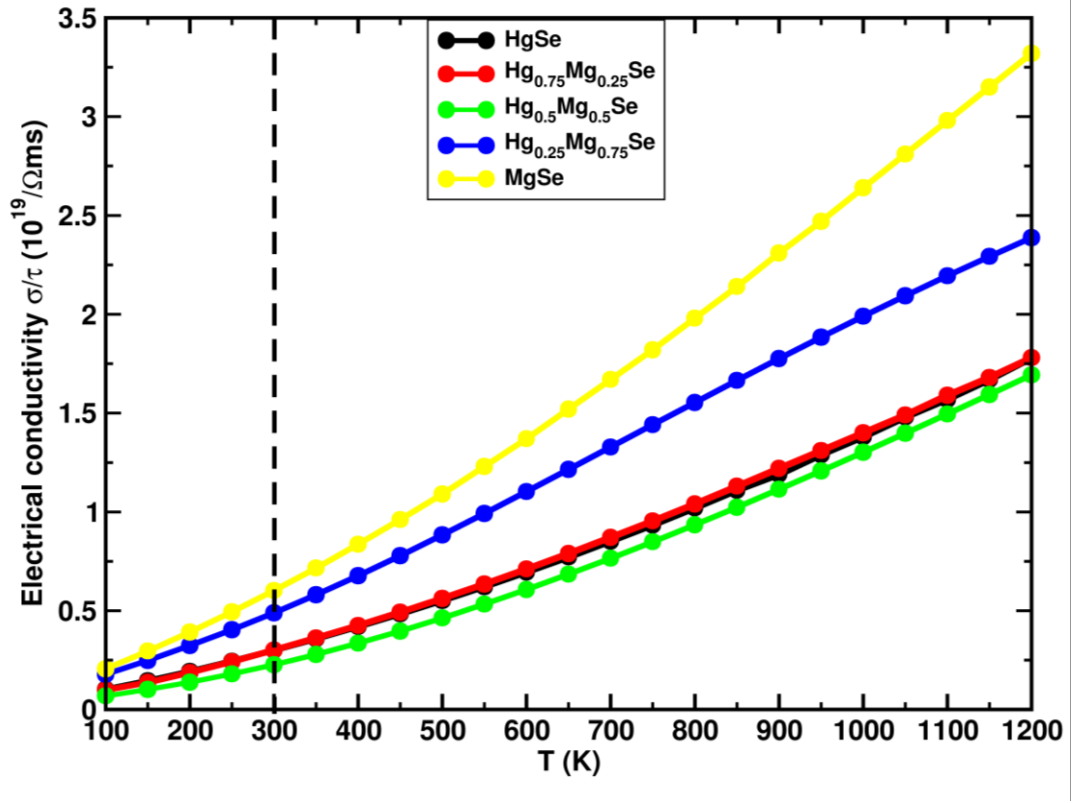

*Figure 5: Calculated electrical conductivity over relaxation time (σ/τ) as a function of temperature (T).*

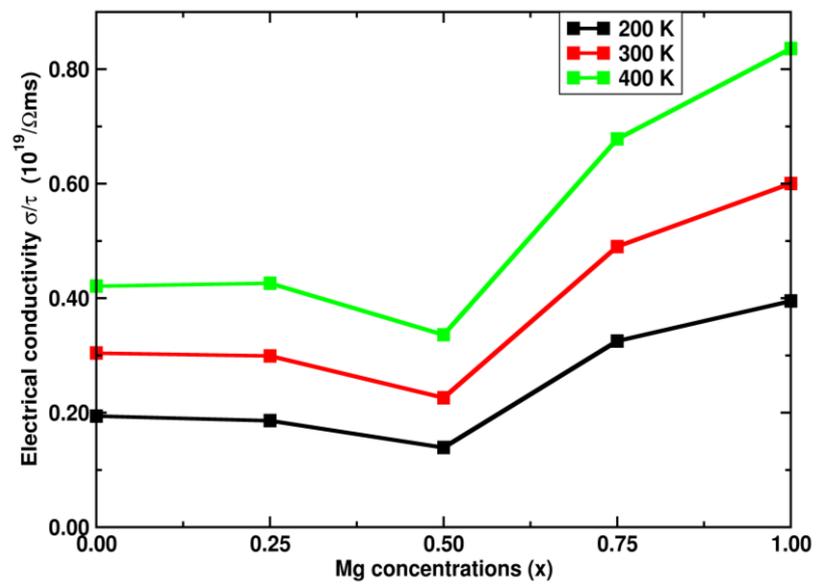

*Figure 6: The calculated electrical conductivity as a function of concentrations of Mg for different temperature.*

## 3.4. Power Factor

The Power Factor **PF** is more important thermoelectric parameters for the thermoelectric power generation, it is calculated using the following equation:

$$PF = S^2\sigma/\tau \qquad (5)$$

Fig. 7 shows the calculated **PF** curves as a function of temperature. At 300 K, all studied materials has a large Power Factors of approximately 0.12, 0.14, 0.12, 0.27 and 0.33 × $10^{12}$ W/mK$^2$s for Hg$_{1-x}$Mg$_x$Se (x=0.0, 0.25, 0.50, 0.75 and 1.0), respectively. Moreover, the power factor increases linearly with the increase in temperature and reaches high values in high temperature region. Due to the enhanced electrical conductivity at room temperature, it is clearly noted that MgSe and Hg$_{0.25}$Mg$_{0.75}$Se has much larger power factor than other studied materials. Additionally, in the temperature range [200-300] K, it is found that with the increase of Mg doping concentration up to 0.50, the power factor decreases. Above the Mg concentration of 0.5, the PF starts to increase to reaches high values, see Fig. 8.

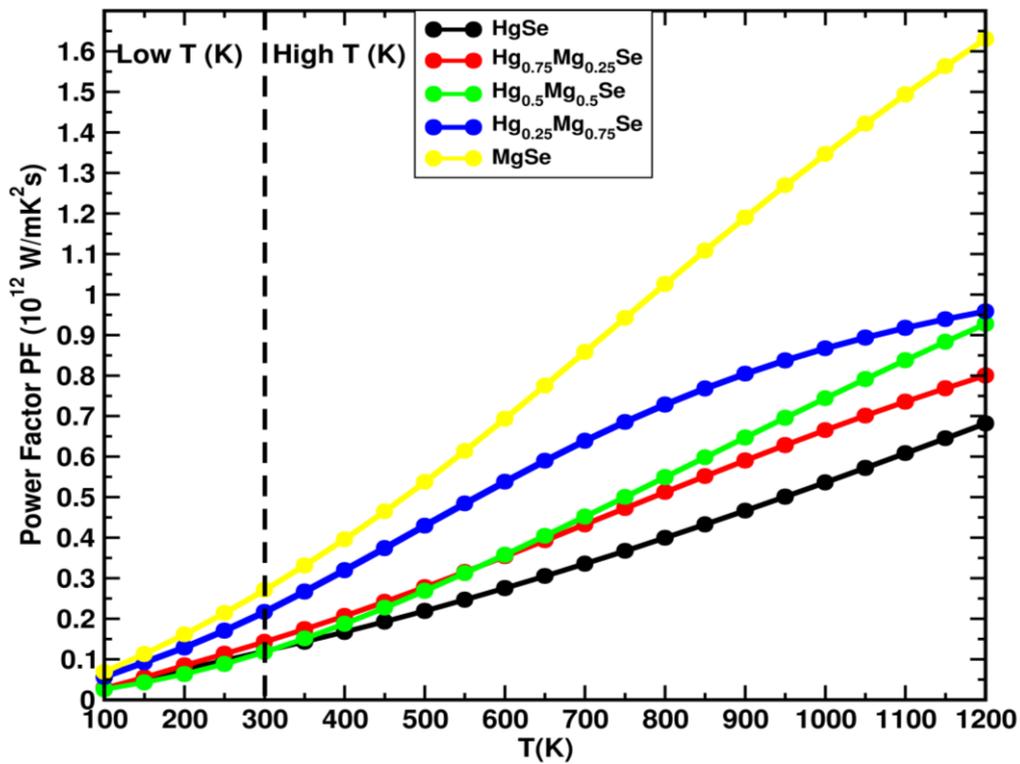

*Figure 7: Calculated Power factor (PF) as a function of temperature (T).*

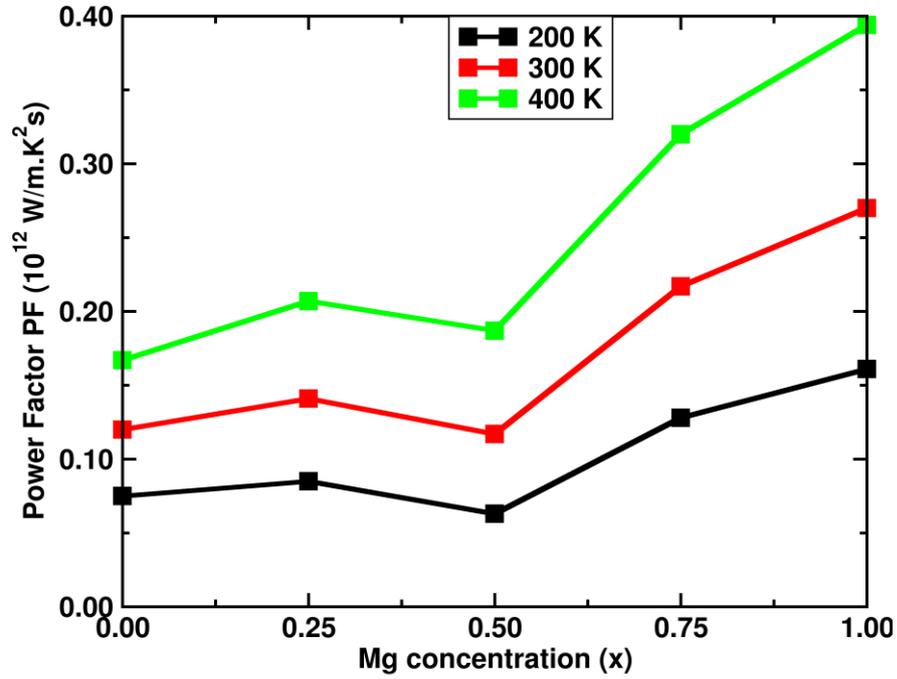

*Figure 8: Power factor as a function of concentrations of Mg for different temperature.*

### 3.5. Figure of merit

The dimensionless Figure of merit (**ZT**) is a transport parameter used to indicate the efficiency of thermoelectric materials. This parameter is totally dependent on the **S, σ, k** and **T**, it can be obtained by the following relation:

$$ZT = \frac{S^2 \sigma T}{k} \qquad (6)$$

The calculated **ZT** values of the studied materials according to the temperature (**T**) are depicted in Fig. 9. It can be seen from this figure that the **ZT** has important values which are near unity. At room temperature (300 K), it is clearly shown that the Figure of merit values are about: 0.62, 0.72, 0.71, 0.70 and 0.71 for $Hg_{1-x}Mg_xSe$ (x=0.0, 0.25, 0.50, 0.75 and 1.0), respectively. These results of **ZT** shows that the $Hg_{0.75}Mg_{0.25}Se$, $Hg_{0.5}Mg_{0.5}Se$, $Hg_{0.25}Mg_{0.75}Se$ and MgSe compounds have important Figure of merit than HgSe. Moreover, in high temperature region we can note that with increasing the doping concentration of Mg, the Figure of merit first increase at x = 0.75, then decreases up to x = 1.0. The calculated results of the Power Factor and the Figure of merit have summarized in the Table. 2.

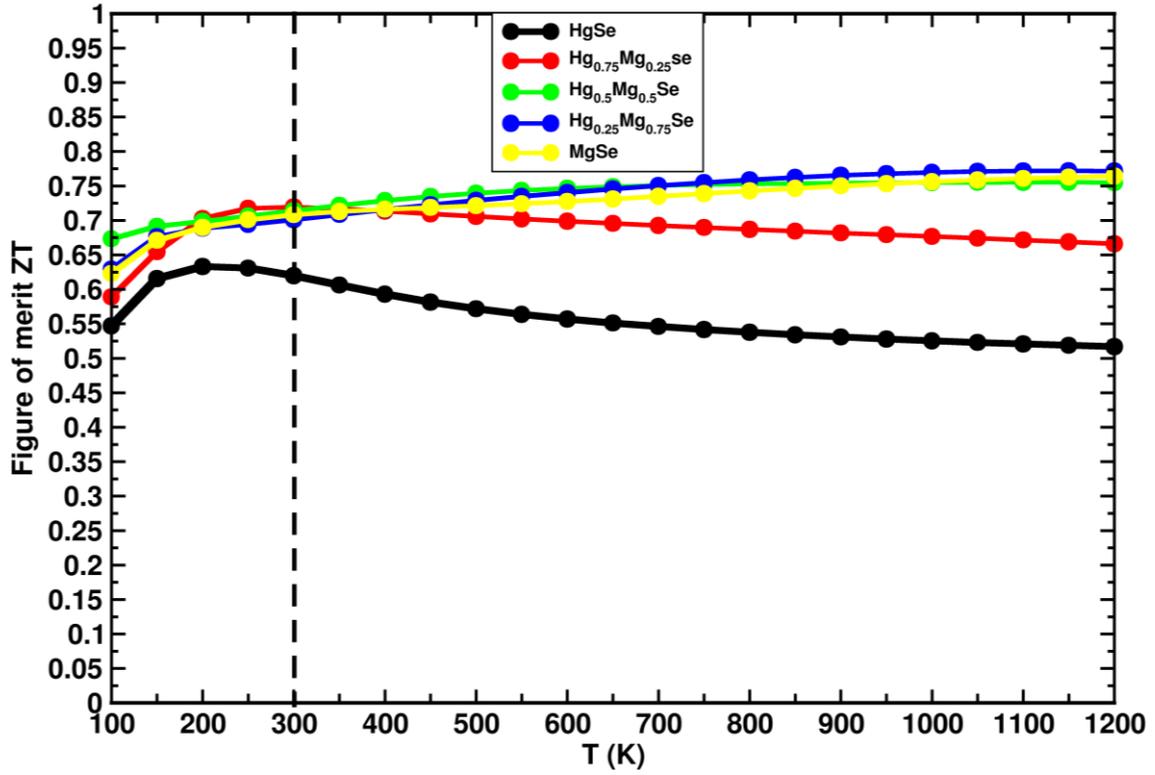

*Figure 9: Calculated electronic Figure of merit (ZT) as a function of temperature (T).*

*Table 1: The calculated Seebeck (**S**), electronic thermal conductivity (**k/τ**) and Electrical conductivity (**σ/τ**) of $Hg_{1-x}Mg_xSe$ (x=0.0, 0.25, 0.50, 0.75 and 1.0) at temperature of 300, 600, 900 and 1200 K.*

|   | Seebeck coefficient S ($10^{-4}$ V/K) | | | | | Electronic thermal conductivity k/τ ($10^{19}$/Ωms) | | | | | Electrical conductivity σ/τ ($10^{14}$ W/mKs) | | | | |
|---|---|---|---|---|---|---|---|---|---|---|---|---|---|---|---|
| x<br>T (K) | 0.0 | 0.25 | 0.50 | 0.75 | 1.0 | 0.0 | 0.25 | 0.50 | 0.75 | 1.0 | 0.0 | 0.25 | 0.50 | 0.75 | 1.0 |
| 300 K | 2.00 | 2.18 | 2.28 | 2.10 | 2.12 | 0.60 | 0.60 | 0.49 | 0.96 | 1.14 | 0.316 | 0.30 | 0.23 | 0.49 | 0.59 |
| 600 K | 1.99 | 2.23 | 2.42 | 2.21 | 2.25 | 2.97 | 3.04 | 2.88 | 4.37 | 5.71 | 0.696 | 0.71 | 0.61 | 1.10 | 1.36 |
| 900 K | 1.98 | 2.20 | 2.41 | 2.13 | 2.27 | 7.92 | 7.84 | 7.73 | 9.47 | 14.29 | 1.19 | 1.22 | 1.11 | 1.78 | 2.30 |
| 1200 K | 1.96 | 2.12 | 2.34 | 2.00 | 2.22 | 15.81 | 14.42 | 14.73 | 14.91 | 15.60 | 1.75 | 1.78 | 1.69 | 2.41 | 3/31 |

*Table 2: The calculated Power Factor (**PF**) and Figure of merit (**ZT**) of $Hg_{1-x}Mg_xSe$ (x=0.0, 0.25, 0.50, 0.75 and 1.0) at temperature of 300, 600, 900 and 1200 K.*

|  | Power Factor $PF$ ($10^{12}$ W/mK$^2$s) | | | | | Figure of merit **ZT** | | | | |
|---|---|---|---|---|---|---|---|---|---|---|
| x<br>T (K) | 0.0 | 0.25 | 0.50 | 0.75 | 1.0 | 0.0 | 0.25 | 0.50 | 0.75 | 1.0 |
| 300 K | 0.12 | 0.14 | 0.12 | 0.27 | 0.33 | 0.62 | 0.72 | 0.71 | 0.70 | 0.71 |
| 600 K | 0.27 | 0.35 | 0.36 | 0.54 | 0.70 | 0.56 | 0.70 | 0.75 | 0.70 | 0.73 |
| 900 K | 0.47 | 0.59 | 0.65 | 0.81 | 1.18 | 0.53 | 0.68 | 0.75 | 0.76 | 0.75 |
| 1200 K | 0.68 | 0.81 | 0.93 | 0.96 | 1.62 | 0.52 | 0.67 | 0.76 | 0.77 | 0.76 |

## 4. Conclusion

In this study, we have illustrated the thermoelectric properties of $Hg_{1-x}Mg_xSe$ with different Mg concentrations (x=0.0, 0.25, 0.50, 0.75 or 1.0). For this purpose, we have applied the first-principles calculations in combination with Boltzmann transport theory. The calculated Seebeck coefficient and thermoelectric properties of $Hg_{1-x}Mg_xSe$ are calculated and analyzed in this study. Our results are performed for such materials at room temperature. Large Seebeck coefficients values, large electrical conductivity values and low electronic thermal conductivity of $Hg_{1-x}Mg_xSe$ (x=0.0, 0.25, 0.50, 0.75 or 1.0) have been obtained and discussed. However, at room temperature, the calculated Figure of merit values of $Hg_{1-x}Mg_xSe$ have been found to be: 0.62, 0.72, 0.71, 0.70 and 0.71. Also, we found that the doping HgSe by Mg compositions increases the figure of merit values. Finally, the thermoelectric results indicate that the $Hg_{1-x}Mg_xSe$ (x= 0.25, 0.50, 0.75 or 1.0) can exhibit promising properties for future thermoelectric applications.